\documentclass[12pt]{article}
\usepackage{amssymb}
\usepackage[dvips]{epsfig}
\usepackage[utf8]{inputenc}
\usepackage{graphicx}
\usepackage{amsmath} 
\usepackage{bbm}
\usepackage{bm}

\usepackage{array}

\usepackage{hyperref}
\usepackage{longtable}

\def \del{\partial}

\newcommand{\be}{\begin{equation}}
\newcommand{\ee}{\end{equation}}
\def\bea{\begin{eqnarray}}
\def\eea{\end{eqnarray}}
\newcommand{\bn}{\begin{eqnarray}}
\newcommand{\en}{\end{eqnarray}}
\newcommand{\nn}{\nonumber}
\newcommand{\no}{\noindent}

\newcommand{\p}{\partial}
\def\bea{\begin{eqnarray}}
\def\eea{\end{eqnarray}}
\newcommand{\beq}{\begin{eqnarray}}
\newcommand{\eeq}{\end{eqnarray}}

\def \eps {\epsilon }
\def \lap {\nabla^2}

\usepackage{indentfirst}

\begin{document}

\title{\textbf{Linearized transverse diffeomorphism invariant spin-2 theories via gauge invariants}}

	\author{D. Dalmazi \footnote{denis.dalmazi@unesp.br}, Luiz G. M. Ramos \footnote{luiz.monteiro@unesp.br}\\
		\textit{{UNESP - Campus de Guaratinguetá - DFI }}} 
	\date{\today}
	\maketitle

\begin{abstract}

We analyze the particle spectrum of a second-order (in derivatives) theory based on a rank-2 tensor field with both symmetric and antisymmetric components. By demanding the existence of a propagating massless spin-2 particle and invariance under linearized transverse diffeomorphisms, we derive a new class of stable models with two massless scalars and a single massless spin-2 particle. A natural nonlinear completion is proposed in terms of a dynamical metric field. The identification of the spectrum is carried out using a fully Lagrangian, gauge-invariant approach which makes use of Bardeen variables in a constructive manner. The approach significantly reduces the number of steps in the spectrum determination in some cases.

\end{abstract}

\newpage

\tableofcontents

\newpage

\section{Introduction}

General relativity (GR) stands as one of the most successful physical theories, yet it remains incomplete. While GR provides a consistent gauge theory for a self-interacting massless spin-2 particle, it faces significant challenges, including non-renormalizability and persistent cosmological tensions such as dark energy, dark matter, the cosmological constant problem, and the Hubble tension. For comprehensive discussions of these issues and alternative theories, we refer to recent reviews \cite{valentino,abdalla}.

From a field-theoretic perspective, modifications to GR can be approached in various ways, see \cite{lh} and references therein. One may, for instance, consider altering the underlying gauge symmetry while maintaining consistency with or without additional degrees of freedom.

As a gauge theory, GR is invariant under general coordinate transformations, which reduce to linearized diffeomorphisms $\delta h_{\mu\nu}=\partial_\mu\epsilon_\nu+\partial_\nu\epsilon_\mu$, when expanded around flat spacetime\footnote{Here, we work with the Minkowski metric $\eta_{\mu\nu} = \mathrm{diag}(-1,1,\dots,1)$ in $D \geq 3$ dimensions, defining $g_{\mu\nu} = \eta_{\mu\nu} + h_{\mu\nu}$.}. However, it is known \cite{Ng} that the minimal symmetry required for a consistent description of massless spin-2 particles is given by transverse diffeomorphisms (TDiff), characterized by $\partial^\mu \epsilon^T_\mu = 0$. These can be extended nonlinearly to volume-preserving diffeomorphisms ($\nabla^\mu \epsilon^T_\mu = 0$). This scalar restriction reduces the number of independent gauge parameters by one, typically introducing an additional scalar degree of freedom alongside the spin-2 mode. Whether this scalar is physical depends on the Lagrangian parameters, as detailed in \cite{blas} and here in section 2.

An exception arises when the model parameters are fine-tuned to accommodate a linearized Weyl symmetry ($\delta h_{\mu\nu} = \sigma \eta_{\mu\nu}$), extending TDiff to WTDiff (Weyl plus TDiff). In this case, the spectrum reduces to only massless spin-2 particles. The WTDiff model coincides with linearized unimodular gravity \cite{hu}. 
It can also be nonlinearly completed to a WTDiff gravity \cite{oda2017}. More generally, TDiff models can also be nonlinearly completed in terms of a metric field, using the fact that the metric determinant transforms as a scalar under volume-preserving diffeomorphisms \cite{blas}.

The TDiff models have some general interesting aspects. Notably, they have an integration constant with the same mass dimension as the cosmological constant which may have interesting phenomenological consequences in scale invariant theories of gravity\footnote{In particular, in TDiff theories there naturally appears an integration constant with the same mass dimension of the cosmological constant which may have relevant application to the dark energy problem, see \cite{bsz}.} \cite{bsz}. In WTDiff gravity, the constant plays the role of a cosmological constant that does not receive quantum corrections, see discussion in \cite{oda2017}. It is also worth commenting that TDiff models
 (with or without Weyl symmetry) cannot become massive in their spin-2 sector \cite{blas}.  Given the stringent experimental bounds on the graviton mass ($m_g \lesssim 10^{-23}~\mathrm{eV}$), the TDiff symmetry seems to be more natural than the usual Diff one (Einstein-Hilbert) which can become massive at linearized level \cite{fp} and beyond \cite{massive,bimetric}, see also \cite{hogas}.
 
 Another active direction in modifying gravity involves metric affine geometries \cite{hehl}, see also the review works \cite{lhf(q)}. There have been intense work specially on flat geometries like  Teleparallel gravities with generalized quadratic torsion terms known as New General Relativity (NGR), see \cite{shira}, quadratic theories on non-metricity with vanishing torsion, \cite{nyo} and \cite{jhk}, and generalized quadratic forms of both non-metricity and torsion \cite{gtg}. In the case of flat geometries with nonvanishing torsion, when expanding about the Minkowski space, either using the vierbein formalism $e_{\mu}^{\,\,\, a} $ or as an element $\Lambda_{\mu}^{\,\,\, \alpha} $ of the group $GL(4,\mathbb{R})$ we have a rank-2 field with symmetric and nonsymmetric parts instead of the usual description of massless spin-2 particles in terms of a symmetric rank-2 field. 
 
 The aim of the present work is twofold, first we merge the previous mentioned directions of modifying gravity at linearized level and investigate the most general Lorentz invariant quadratic theory of second order in derivatives invariant under TDiff and described by a full rank-2 tensor. Thus, paving the way for non linear completions. Secondly, the content of the model  is examined by using an explicitly gauge invariant approach based on an improved use of Bardeen variables \cite{Jaccard} which turns out to be a very convenient way of unravelling the particle content of quadratic tensor models.
 
 The paper is structured as follows. In section 2 we introduce the method by applying it to simpler examples. In section 3 we present the general models we are interested in and explore their particle content. A class of models with one massless spin-2 particle and two  scalars, which we call a SST (Scalar-Scalar-Tensor) model, is coupled to external sources in section 4 where we also suggest its natural nonlinear completion. In section 5 we draw our conclusions and outline future directions. The appendix provides an alternative derivation of the spectrum of the free SST model via the introduction of an auxiliary field.

\section{TDiff model via Bardeen variables}

\subsection{TDiff spectrum via a shortcut}

The most general massless TDiff invariant local spin-2 model of second order in derivatives, described only in terms of a symmetric rank-2 field, is given by the following Lagrangian:

\be \mathcal L_{TD}(a,b) = \frac{1}{2} h^{\alpha\beta}\, \Box \,  h_{\alpha\beta}+(\del^\mu h_{\mu\beta})^2+ a\, h\, \del^{\mu}\del^{\nu} h_{\mu\nu}-\frac{b}{2} h\,  \Box h \, .
\label{sab}
\ee

\no The first two terms are required for massless spin-2 propagation without
spin-1 ghosts \cite{blas}. The last two terms depend on two real constants: $(a,b)$. Up to surface terms, neglected throughout this work, the lagrangian in (\ref{sab}) is TDiff invariant\footnote{ Here we use the index ``$T$'' for space-time ($\p^{\mu}v_{\mu}^T$=0) and ``$t$'' for space ($\p_{j}v_{j}^t=0$) transverse quantities.} ($ \del^\mu \eps^T_\mu=0$) for arbitrary $(a,b)$,
\be 
\delta h_{\mu\nu} = \del_\mu \eps^{T}_{\nu} +  \del_\nu \eps^{T}_{\mu}\, . \label{tdiffs} \ee

\no The content has been previously investigated in \cite{blas}  also via gauge invariants but differently from our simpler approach as we will show here, in \cite{rrm} via the usual Hamiltonian Dirac procedure for constrained systems and in \cite{rr} via the analytic structure of the propagator.  There are two points in the parameters space  where TDiff is enhanced \cite{blas}, 

\bea  \delta h_{\mu\nu} &=& \del_\mu \eps_{\nu} +  \del_\nu \eps_{\mu} \,  \quad ; \quad (a,b)=(1,1) \label{diff}, \\  \delta h_{\mu\nu} &=& \del_\mu \eps^{T}_{\nu} +  \del_\nu \eps^{T}_{\mu} + \eta_{\mu\nu} \, \sigma  \quad ; \quad (a,b)=  \left(\frac 2D,\frac{D+2}{D^2}\right) \equiv (a_D,b_D).  \label{wtd}  \eea

\no In the first case, we have the linearized Einstein-Hilbert\footnote{More precisely, there is a continuous family of models  $(a(r),b(r))$ equivalent to the point $(a,b)=(1,1)$ via what we call r-shifts: $h_{\mu\nu} \to h_{\mu\nu} + r\, \eta_{\mu\nu} \, h$ for which the enhancement still holds, though (\ref{diff}) is  modified accordingly, while the case $(a,b)=(a_D,b_D)$ is a fixed point of the r-shifts \cite{blas}.}  theory $\mathcal{L}_{TD}(1,1)=\mathcal{L}_{EH}$ where TDiff is enhanced to  Diff. In the second one, the symmetry becomes Weyl + TDiff  (WTDiff). In both cases, the model describes only massless spin-2 particles  in $D=4$ with only two degrees of freedom corresponding to the helicities $\pm 2$. 

For an otherwise arbitrary couple $(a,b)$, a quite lengthy analysis, see \cite{blas},\cite{rrm}  and  \cite{rr}, reveals that we have in general $(D-1)(D-2)/2$ degrees of freedom corresponding to a massless spin-2 and a massless spin-0  particle in $D=4$. A key quantity is $f_D(a,b)$:

\be f_D(a,b) \equiv (a^2-b)(D-2) + (a-1)^2 \quad . \label{fd} \ee

\no If $f_D(a,b)=0$ the spin-0 particle disappears, which is the case of (\ref{diff}) and (\ref{wtd}). If $f_D > 0$ ($f_D <0$) the spin-0 particle is physical (ghostly). Remarkably,  we can arrive at the same conclusion of \cite{blas,rrm,rr} in just one step. An important ingredient is that a Lorentz covariant free gauge theory can be written as a quadratic form on gauge invariants ($I_A$) which, on their turn, are linear on the gauge field potentials and their derivatives, symbolically,

\be {\cal L} = I_A \, {\cal M}_{AB} \, I_B \quad , \label{matrix} \ee

\no where $A,B=0, \cdots, N_I$. The matrix ${\cal M}_{AB}$ depends on space and time derivatives. The invariant indices $A,B$ are just labeling indices. They are deprived of meaning regarding Lorentz transformations. Regarding their range, $ N_I$ it is the number of independent gauge invariants.
We will show that $N_I$ equals the number of fields minus the number of independent gauge parameters, generically,

\be N_I = N({\Phi}) - N(\eps)  \quad . \label{ni}\ee

  When we go from Diff to TDiff
we have one less independent gauge parameter due to $\p^{\mu}\eps_{\mu}^T=0$, consequently, from (\ref{ni}) we must have one extra gauge invariant. Since any Diff invariant must be also TDiff invariant and the trace $h$ is TDiff invariant by itself, we can consider   $I_{\tilde{A}}^{TDiff}=\left\lbrace I_A^{Diff},h\right\rbrace$ as a basis of independent TDiff invariants. The trace $h$ cannot be a function of the Diff invariants $I_A^{Diff}$, otherwise it would be also Diff invariant which is certainly not the case, recall $\delta\, h = 2\, \p \cdot \eps\ne 0$. So if we were able to rewrite (\ref{sab}) as a sum of the linearized EH theory and a term that depends only on the trace $h$ we could immediately identify the particle content of (\ref{sab}). Indeed, there is a unique field redefinition which accomplishes it, namely, after $h_{\mu\nu} \to h_{\mu\nu} + r\, \eta_{\mu\nu} h$ with $r=(1-a)/(a D-2)$, assuming $a\ne 2/D$, the TDiff theory (\ref{sab}) can be written, in terms of the new field, as 

\be {\cal L}_{TD}(a,b) = {\cal L}_{EH}(h_{\mu\nu}) + \frac{(D-2)\, f_D}{2(a\, D-2)^2} \, h\, \Box \, h = I_A^{Diff}{\cal M}_{AB}^{EH} \, I_B^{Diff}  + \frac{(D-2)\, f_D}{2(a\, D-2)^2} \, h\, \Box \, h \, , \label{heh} \ee

\no where $I_A^{Diff}{\cal M}_{AB}^{EH}\, I_B^{Diff}$ will be given explicitly later on, see first line of (\ref{ssd}).

Thus, although the field $h_{\mu\nu}$ in ${\cal L}_{EH}(h_{\mu\nu})$ is not traceless, it turns out that the particle content of  ${\cal L}_{TD}(a,b)$ can be read off from the two independent terms in (\ref{heh}),  in full agreement with the more laborious analysis of \cite{blas}, \cite{rrm} and \cite{rr}. Namely, it corresponds to a physical massless spin-2 particle plus a massless scalar particle which is physical (ghostly) if $f_D(a,b)>0$ ($f_D(a,b)<0$) and disappears at $f_D(a,b)=0$.

 In the singular case $a=a_D=2/D$ we are not allowed to perform the ``r-shift'' in the fundamental field $h_{\mu\nu}$. Fortunately, it is not necessary. The reader can check that

 \be {\cal L}_{TD}(a_D,b) = {\cal L}_{WTD}(h_{\mu\nu})   + \frac{ f_D(a_D,b)}{2(D-2)} \, h\, \Box \, h =  {\cal L}_{WTD}(I_A^{WTD})  + \frac{ f_D(a_D,b)}{2(D-2)} \, h\, \Box \, h \, . \label{hwtd} \ee

\no where ${\cal L}_{WTD}(h_{\mu\nu}) ={\cal L}_{TD}(a_D,b_D)$, with $f_D(a_D,b) = (D-2)(b_D-b)$ and $I_A^{WTD}$ are the invariants under WTDiff. Notice that $N_I^{WTD}=N_I^{Diff}$ since the longitudinal component of $\eps_{\mu}$ is replaced by the Weyl gauge parameter such that the number of independent gauge parameters remains the same. On the other hand, as  we go from WTDiff to TDiff, the broken Weyl symmetry gives rise to one extra invariant which may be  again the trace $h$ that now breaks the Weyl symmetry such that 
we can choose  $I_{\tilde{A}}^{TDiff}=\left\lbrace I_A^{WTD},h\right\rbrace$ as another basis for the TDiff invariants. The physical content of ${\cal L}_{TD}(a_D,b)$ factorizes into the content of the two terms in (\ref{hwtd}), once again in agreement with \cite{blas}, \cite{rrm} and \cite{rr}.

Our gauge invariant approach tacitly assumes that Lorentz invariant free gauge models can be consistently described by a reduced number of independent gauge invariants instead of the original field potential $h_{\mu\nu}$.  In the next subsection, we built up the corresponding gauge invariants constructively for each symmetry class and show their independence explicitly. 
The aim of the present section is to explain the technique that will be used in section 3 in the identification of new models.

\subsection{Diff, WTDiff and TDiff gauge invariants}

As carried out in \cite{da-hs} for higher spin analogues of linearized topologically massive gravity and New Massive Gravity in $D=2+1$, here we built up the gauge invariants of a free theory constructively from the definition of the gauge transformations. We begin with the linearized Diff invariants. The invariants can be built up from the Diff gauge transformations without referring to any specific action. From   $D$ out of the $D(D+1)/2$ equations:

\be \delta h_{\mu\nu}=\partial_\mu\epsilon_\nu+\partial_\nu\epsilon_\mu\label{liberdadeDIFF},\ee 

\no we can express $\epsilon_\mu=\epsilon_\mu(\delta h_{\mu\nu})$. It is convenient to decompose the gauge parameters in terms of $SO(D-1)$ scalars and  vectors,
\be \epsilon_0=\alpha\ , \qquad \epsilon_j=\beta_j^t+\partial_j\lambda\, \label{decome}.\ee 

\no  Substituting back in (\ref{liberdadeDIFF}) we obtain the $D$ gauge parameters\footnote{We assume vanishing fields at infinity, so $\nabla^2 \Phi \equiv \p_j^2 \Phi=0  $ leads to $\Phi =0$, where $\Phi$  represents any spacetime field or its spacetime derivatives. Throughout this work the expression $(1/\nabla^2)$  stands for the corresponding Green's function, for instance, in $D=4$ we have $(1/\lap)f (\vec{x},t)= \int\, d^3r^{\prime} f(\vec{x}^{\prime},t)/(4\pi\vert r-r^{\prime}\vert)$.} $\eps_{\mu}(\delta\, h_{\mu\nu})$ :

\be\alpha=\delta\left[\frac{\partial_j h_{0j}}{\nabla^2}-\frac{\omega_{ij} \dot{h}_{ij}}{2\nabla^2}\right]\ ; \qquad \beta^{t}_i=\delta\left[\frac{\theta_{ij}\partial_k h_{jk}}{\nabla^2}\right] \ ; \qquad \lambda=\delta\left[\frac{\omega_{ij} {h}_{ij}}{2\nabla^2}\right],\label{helicitys}\ee 

\no where $\omega_{ij}=\partial_i\partial_j/\nabla^2$ and $\theta_{ij}=\delta_{ij}-\omega_{ij}$ are longitudinal and transverse space projection operators. Using (\ref{helicitys}) into (\ref{liberdadeDIFF}), the remaining $D(D+1)/2-D$ equations become invariance equations $\delta I_A^{Diff}=0$, where $I_A^{Diff}=\{I_{ij}^{t},I_{j}^{t},I_{00}\}$ are the Diff invariants:
\be	I_{00} =h_{00}+\frac{\omega_{ij}\ddot{h}_{ij}}{\nabla^2}-2\frac{\partial_j \dot{h}_{0j}}{\nabla^2}\ ;\qquad  I_{i}^t=\theta_{ij}\left[h_{0j}-\frac{\partial_{k}\dot{h}_{jk}}{\nabla^2}\right]\ ;\qquad I^{t}_{ij}= \theta_{ik}\theta_{jl}h_{kl}\ .\ee

\no They represent a total of $N_I^{Diff}=1+(D-2)+[(D-1)D/2-(D-1)]=(D-1)D/2$ independent Diff invariants in agreement with (\ref{ni}): $N(h_{\mu\nu}) - N(\eps_{\mu})=D(D+1)/2-D$. It is convenient to redefine them on physical grounds. Since the two helicity states $\pm 2$ in $D=4$ correspond to the  traceless and transverse components of $h_{ij}$, we decompose $I_{ij}^{t}$ into its traceless ($I_{ij}^{tt}$) and trace ($I$) components. Respectively,

\bea 	I^{tt}_{ij}&=&I^t_{ij}-\frac{\theta_{ij}}{(D-2)}\delta_{kl}I_{kl}^t\quad , \label{itt} \\  I&=&\delta_{ij}I^t_{ij}=\theta_{kl}h_{kl} \label{I}.\eea

\no Both $\{I^{tt}_{ij},I_{i}^t\},$ are also invariant under Weyl transformations 

\be \delta_w h_{\mu\nu}=\eta_{\mu\nu}\sigma .\label{Weyl2}\ee

\no It will be convenient to have the largest possible number of Diff invariants which are also Weyl invariants. It is easy to check that the combination $I_W\equiv (D-2)\, I_{00} + \Box\, I/\nabla^2$ is both Diff and Weyl invariant. Henceforth we choose to work with the following basis of Diff invariants:

\be I_A^{Diff} = \{I^{tt}_{ij},I_{i}^t,I_W,I\} \quad . \label{diffinv} \ee

Regarding the WTDiff invariants, they can be easily derived from the Diff invariants.
First, we recall that  the number of WTDiff invariants equals the number of Diff invariants. Moreover, under WTDiff transformations, the traceless combination $\bar{h}_{\mu\nu}(h) = h_{\mu\nu} - \eta_{\mu\nu}\, h/D$ changes as $\delta \bar{h}_{\mu\nu} = \p_{\mu} \eps^T_{\nu} + \p_{\nu} \eps^T_{\mu}$, therefore all Diff invariants $I_A^{Diff}(\bar{h}_{\mu\nu}(h))$ are automatically WTDiff invariants. We can choose $I_A^{WTD}(h) = I_A^{Diff}(\bar{h}(h))$. Thus, due to the Weyl invariance of $\{I^{tt}_{ij},I_{i}^t,I_W\} $, we can use the following basis of WTDiff invariants, 

\be I_A^{WTDiff} = \{I^{tt}_{ij},I_{i}^t,I_W,I_{WT}\} \quad . \label{wtdiffinv} \ee

\no where $I_{WT}= I[h(\bar{h})] = \theta_{ij}h_{ij}-(D-2)\,h/D $.

Moving now to the TDiff gauge symmetry, defined by (\ref{tdiffs}), replacing (\ref{helicitys}) in  $\p^{\rho}\eps_{\rho}(\delta\, h_{\mu\nu})=0$ we obtain
another invariant $\delta \, (I_{00} + h - I)=0$, since $\{I_{00},I\}$ already belong to the Diff basis, the new invariant is identified with the trace $h$. From now on we use the following basis of TDiff invariants:

\be I_A^{TDiff} = \{I^{tt}_{ij},I_{i}^t,I_W,I,h\} \quad . \label{tdiffinv} \ee

Now we show that all those gauge invariants can be treated as independent field coordinates. Let us introduce an invertible decomposition, also called helicity decomposition, similar to \cite{blas} and \cite{Jaccard}, see also the much earlier work \cite{djt},
\be h_{00}={\rho};\ \	h_{0j}=\gamma^t_j+\partial_j\Gamma; \ \ h_{ij}=\Psi_{ij}^{tt}+\partial_i \phi^t_j+\partial_j \phi^t_i+\nabla^2(\theta_{ij}\Lambda+\omega_{ij}\psi)
\label{decomph},\ee 

\no The TDiff invariants become,
\be I^{tt}_{ij}=\Psi_{ij}^{tt}\ ;\qquad I^t_{j}= \gamma_j^t-\dot\phi^t_j\ ;\qquad I_W=(D-2)[\rho+\Box\Lambda+\ddot{\psi}-2\dot \Gamma]\;\nn\ee
\be I=(D-2)\nabla^2\Lambda\ ;\qquad h=-\rho+\nabla^2(D-2)\Lambda+\nabla^2\psi\ \label{invariantes}\ee 

\no We can always associate $I_{ij}^{tt}$ and $I_{j}^t$ with the field components $\Psi_{ij}^{tt}$ and ${\gamma_j^t}$ respectively, their independence is clear.  The independence is less trivial in the $SO(D-1)$ scalar sector involving $\{I_W,I,h\}$. At first sight we might associate those three scalars with $\{\rho,\Lambda,\psi\}$ respectively, however when we get $\rho(I_W,I,\psi,\Gamma)$ from  $I_W$ and $I$ and plug back in the trace $h$, all helicities will appear under time derivative. In order to avoid it we must first get rid of $\ddot{\psi}$ inside $I_W$ by redefining $\Gamma\equiv \bar{\Gamma} + \dot{\psi}/2$. So we end up with a change of variables from 
$\{\rho,\Lambda,\psi,\Gamma\}$ to $\{I_W,I,h,\bar{\Gamma}\}$. We can check that the Jacobian does not depend upon time derivatives, namely,

\be \begin{bmatrix}
 I_W \\
 I \\
 h \\
 \bar\Gamma
 \end{bmatrix}=\begin{bmatrix}
 (D-2) & (D-2)\Box & (D-2)\del_0^2 & -2(D-2)\del_0 \\
 0 & (D-2)\nabla^2 & 0 & 0 \\
  -1 & (D-2)\nabla^2 & \nabla^2 & 0 \\
 0 & 0 & -\del_0/2 & 1 
 \end{bmatrix}\begin{bmatrix}
 \rho \\
 \Lambda \\
 \psi \\
 \Gamma
 \end{bmatrix}\equiv\mathbb M_{TD}\begin{bmatrix}
 \rho \\
 \Lambda \\
 \psi \\
 \Gamma
 \end{bmatrix}\label{transma}\ee where, despite the time derivatives, the matrix  $\mathbb M_{TD}$ is non-singular according to our boundary conditions, i.e., \be det[\mathbb M_{TD}]=(D-2)^2\nabla^4,\ee The coordinates transformation is invertible and the gauge invariants are linearly independent as the helicity variables. We conclude that indeed the particle content of the two terms in (\ref{heh}) can be read off separately.

We can show as well that the $SO(D-1)$ scalar sector of $I_{A}^{WTD}$ (i.e. $\{I_W,I_{WT},h\}$)  form an independent basis. It turns out that  $\det[\mathbb M^{(W)}_{TD}]=\det[\mathbb M_{TD}]=(D-2)^2\nabla^4$, where $ M^{(W)}_{TD}$ is the matrix appearing in the change of coordinates from $\{\rho,\Lambda,\psi,\Gamma\}$ to $\{I_W,I_{WT},h,\bar{\Gamma}\}$, where $ I_{WT}=(D-2)(\rho+2\nabla^2\Lambda-\nabla^2\psi)/D $.

\section{NSTDiff models via Bardeen variables}

\subsection{NSTDiff spectrum via a shortcut}

In the present section we extend the theories presented in the last section by replacing the symmetric rank-2 fundamental field by a general rank-2 tensor with both symmetric $(h_{\mu\nu}=h_{\nu\mu})$ and antisymmetric ($B_{\mu\nu}=-B_{\nu\mu}$) parts: $e_{\mu\nu} = h_{\mu\nu} + B_{\mu\nu}$. 
Starting from a general theory of second order in derivatives and requiring symmetry under nonsymmetric TDiff (NSTDiff):  $\delta e_{\mu\nu} = \p_{\nu}\,\eps_{\mu}^T$ we end up with  two possibilities. They both depend in principle on three arbitrary real parameters\footnote{We could have defined $ \delta e_{\mu\nu} = a\, \p_{\mu}\eps^T_{\nu} + (1-a) \p_{\nu} \eps^T_{\mu} $ with some arbitrary $a$, however this could be undone by  redefining the fundamental field taking combinations of $e_{\mu\nu}$ and $e_{\nu\mu}$.}. 

In the first case $B_{\mu\nu}$ is decoupled from $h_{\mu\nu}$,

\be \tilde{{\cal L}}_{NST}(a,b,\tilde{c})  = \frac{1}{2} h^{\alpha\beta}\, \Box \,  h_{\alpha\beta}+(\del^\mu h_{\mu\beta})^2+ a\, h\, \del^{\mu}\del^{\nu} h_{\mu\nu}-\frac{b}{2} h\,  \Box h  + \, \tilde{c}\, [ B_{\mu\nu} \, \frac{\Box}2 \, B^{\mu\nu}+ ( \p^{\mu}B_{\mu\nu})^2 ] .
\label{labct}
\ee

\no where $\tilde{c}$ is a real constant. The new model (\ref{labct}) is invariant under

\be \delta h_{\mu\nu} = \del_{\nu}\,\epsilon_{\mu}^T + \del_{\mu}\,\epsilon_{\nu}^T  \quad ; \quad \delta B_{\mu\nu} = \del_{\nu}\,\lambda^T_{\mu}- \del_{\mu}\,\lambda^T_{\nu}\quad , \label{deltatil} \ee

\no where $\lambda_{\mu}^T$ and $\eps_{\mu}^T$ are independent parameters. The content of (\ref{labct}) is of course the content of (\ref{sab}) added to the content of the two-form field ($B_{\mu\nu}$) Lagrangian, namely a massless scalar tensor theory where the scalar is physical if $f_{D}(a,b)>0$, plus, in $D=4$, a physical  pseudo scalar if $\tilde{c} >0$. At the point $(a,b)=(1,1)$ the model (\ref{labct}) is just the sum of the LEH theory (massless spin-2) and the pseudo scalar coming from the two form, such model is the linearized version of the so called one-parameter teleparallel gravity or one parameter NGR (new general relativity) \cite{ccn}. In $D=4$ we find a dual equivalent model by replacing $B_{\mu\nu} \to \epsilon_{\mu\nu\alpha\beta}\tilde{B}^{\alpha\beta}$.

In the second case $B_{\mu\nu}$ is coupled to $h_{\mu\nu}$,

\be {\cal L}_{NST}(a,b,c)  = \frac{1}{2} h^{\alpha\beta}\, \Box \,  h_{\alpha\beta}+(\del^\mu h_{\mu\beta})^2+ a\, h\, \del^{\mu}\del^{\nu} h_{\mu\nu}-\frac{b}{2} h\,  \Box h \, + \, c\, ( \p^{\mu}e_{\mu\nu})^2 \quad ,
\label{labc}
\ee

\no where $c$ is a real constant. It turns out that (\ref{labc}) is invariant under:

\be \delta e_{\mu\nu} = \p_{\nu}\,\epsilon_{\mu}^T + \Lambda_{\mu\nu}^T \qquad {\rm (NSTDiff)} \quad , \label{dnstd} \ee

\no where  $\Lambda_{\mu\nu}^T=-\Lambda_{\nu\mu}^T$ and $\p^{\mu}\Lambda_{\mu\nu}^T=0$. 

Similarly to the previous section, there are two cases where NSTDiff is enhanced,

\bea  \delta e_{\mu\nu} = \p_{\nu}\,\eps_{\mu} + \Lambda_{\mu\nu}^T \,  \quad &;& \quad a=b=2\, c+1 \qquad ; \qquad {\rm (NSDiff)} \, , \label{nsd} \\   \delta e_{\mu\nu}  = \p_{\nu}\,\eps_{\mu}^T + \Lambda_{\mu\nu}^T + \eta_{\mu\nu} \, \sigma  \quad &;& \quad (a,b)= [a_D(c),b_D(c)]  \quad ; \quad  {\rm (NSWTDiff)} \, , \label{nswtd}   \eea

\no where \be a_D(c) \equiv 2(1+c)/D\ , \qquad b_D(c) \equiv \frac{[D+2(1+c)]}{D^2}\ .\ee We recover (\ref{diff}) and (\ref{wtd}) at $c=0$.

The question arises: what is the particle content of ${\cal L}_{NST}(a,b,c)$ ? We follow now the same shortcut of the symmetric case. We search for an r-shift $e_{\mu\nu} \to e_{\mu\nu} + r\, \eta_{\mu\nu}\, h$  which might split ${\cal L}_{NST}(a,b,c)$ into two pieces whose content could be investigated separately. First, assuming that $a\ne a_D(c)$, we can always choose $r=(2c+1-a)/[a\, D -2(1+c)]$
 which leads to a sum of a general NSDiff model  plus a kinetic term for the trace $h$. Namely,

     \bea {\cal L}_{NST}(c,B) &=&   h^{\alpha\beta} \frac{\Box}{2} h_{\alpha\beta}+(\del^\mu h_{\mu\beta})^2+ (2\, c+1)\left( h\, \del^{\mu}\del^{\nu} h_{\mu\nu}- h\,  \frac{\Box}{2} h \right)  +  c\, ( \p^{\mu}e_{\mu\nu})^2 + \frac{B}2 h\, \Box \, h \nn\\ &\equiv & {\cal L}_{NSD}(c) + \frac{B}2 h\, \Box \, h \label{lc}
\eea
 
 \no where the constant $B$ is an analytic function of $(a,b,c)$, except at 
 $a=a_D(c)$, whose details are not important for us as we explain below. The model ${\cal L}_{NSD}(c)$ defined in (\ref{lc}) is   invariant under (\ref{nsd}). It has first appeared\footnote{The model ${\cal L}_{NSD}(c)$ corresponds to the ${\cal L}(a_1)$ theory in \cite{rank-2} with the map $c=a_1-1/4$.} in  \cite{rank-2} where one has searched for the most general Lorentz invariant massive spin-2 model of second order in derivatives of an arbitrary rank-2 tensor. Those requirements have led to a family of massive models given in terms of an arbitrary real parameter, see \cite{rank-2}, which includes the paradigmatic Fierz-Pauli model \cite{fp}. Its massless limit is given by ${\cal L}_{NSD}(c)$.
 
 Regarding the case $a=a_D(c)=2(1+c)/D$, we can, similarly to the symmetric case, split ${\cal L}_{NST}[a_D(c),b,c]$ into a NSWTDiff model plus a kinetic term for the trace without need of an r-shift,
 
 \be {\cal L}_{NST}[a_D(c),b,c] = {\cal L}_{NST}[a_D(c),b_D(c),c] + \frac{\tilde{B}}2 \, h\, \Box \, h \equiv {\cal L}_{NSWTD}(c) + \frac{\tilde{B}}2 \, h\, \Box \, h \quad , \label{lnswtd} \ee 
 
 \no where $\tilde{B} = b_D(c)-b$.
 
 Since the general NSTDiff model (\ref{labc}) can either be brought to the form (\ref{lc}) or (\ref{lnswtd}), henceforth we refer to the latter models as our definition of a general NSTDiff model without loss of generality and the r-shift has reduced the  effective number of arbitrary constants to only two: $(c,B)$. Next we investigate the particle content of (\ref{lc}) and (\ref{lnswtd}). 
 
We will show in the next subsection that ${\cal L}_{NSD}(c)$ can be written in terms of NSDiff gauge invariants $I_A^{NSD}$,   $A=1,2, \cdots, N_I^{NSD}$ where $N_I^{NSD} = D^2-[D+D(D-1)/2-(D-1)]$. The transverse condition $\p^{\mu}\eps_{\mu}^T=0$ reduces one unit of gauge parameters which leads to one extra gauge invariant which we chose to be the trace $h$ again. Thus, we will use the following basis for the NSTDiff invariants $I_{\tilde{A}}^{NSTD}=\left\lbrace I_A^{NSD} , h \right\rbrace $. In order to find out the content of (\ref{lc}) we only need to know the content of ${\cal L}_{NSD}(c)$. Its content has been investigated  via a Lagrangian method based on the introduction of an auxiliary vector field and confirmed in \cite{bds} via the usual hamiltonian Dirac method. The model ${\cal L}_{NSD}(c)$ has been rediscovered recently in \cite{bhbd}, where it is named type 3, as a special case of linearized NGR about Minkowski space. In the next subsection, we show via gauge invariants that it describes, see also \cite{rank-2}, a physical massless spin-2 particle and a spin-0 particle which is physical if either $c>0$ or $c <- c_D$ where 

\be c_D =  \frac{(D-2)}{2(D-1)}\ .\ee 

\no At the border cases $c=0$ and $ c=-c_D$ the spin-0 particle disappears. At $c=0$ the transverse condition on the antisymmetric shifts is lifted ($\Lambda_{\mu\nu}^T \to \Lambda_{\mu\nu}$) and we can gauge away $B_{\mu\nu}$ and arriving at the usual EH model. At $c=-c_D$ there appears a Weyl symmetry ($\delta e_{\mu\nu} = \sigma\, \eta_{\mu\nu}$) so we have a  NSWDiff model, previously found even earlier in \cite{cmu} and also rediscovered in \cite{bhbd} where it is called type 8.  So regarding (\ref{lc}) we conclude that we have a massless spin-2 particle and two physical spin-0 particles if $B>0$ and either $c>0$ or $c<-c_D$.

 Regarding the special case $a=a_D(c)$ in (\ref{lnswtd}), we have exactly the same content of (\ref{lc})  under the same conditions, i.e., a physical massless SST theory whenever $B>0$ and either $c>0$ or $c<-c_D$. The coincidence stems from the fact that, as in the symmetric case of the last section, we have ${\cal L}_{NSWTD}(h_{\mu\nu}) = {\cal L}_{NSD}[\bar{h}_{\mu\nu}(h)]$, likewise for the invariants $I_A^{NSWTD}(h_{\mu\nu}) = I_A^{NSD}[\bar{h}_{\mu\nu}(h)]$. Since they are all independent quantities as we will show here, the same spectrum must hold under the same conditions. In table 1 we summarize the linearized new models 
 suggested here. 
 
  \begin{table}[h]
		\centering
		\begin{tabular}{l|l|l|l} 
			\hline
	{ Model }	& { Formula }  & { Symmetries } & { Restrictions }   \\
			\hline
			\hline
			$\tilde{{\cal L}}_{NST}(a,b,\tilde{c}) $ & \quad (24)	& $\delta h_{\mu\nu} = \del_{\nu}\,\epsilon_{\mu}^T +
 \del_{\mu}\,\epsilon_{\nu}^T   $                                                                    & $f_D(a,b) >0$ \, and \, $\tilde{c}>0$ \\
 & &  $\delta B_{\mu\nu} = \del_{\nu}\,\lambda^T_{\mu}- \del_{\mu}\,\lambda^T_{\nu}$ &  \\ \hline
 ${\cal L}_{NST}(c,B)$	&   \quad (31) & $\delta e_{\mu\nu} = \p_{\nu}\,\epsilon_{\mu}^T + \Lambda_{\mu\nu}^T $ & $c>0$ \, or \, $c<-c_D$ and $B>0$ \\ \hline
  ${\cal L}_{NST}(a_D,b,c)$	&  (32) \, and \, (26) & $\delta e_{\mu\nu}  = \p_{\nu}\,\eps_{\mu}^T + \Lambda_{\mu\nu}^T  $ & $c>0$ \, or \, $c<-c_D$ and $\tilde{B}>0$ \\ \hline \hline
		\end{tabular}
		\caption{New models with their local symmetries and restrictions on the parameter's space. They all have the same particle content, one spin-2 and two spin-0 massless modes (all physical). Notice that $(h_{\mu\nu},B_{\mu\nu}) = (e_{(\mu\nu)},e_{[\mu\nu]}) $}
	\end{table}

\subsection{NSDiff, NSWTDiff  and NSTDiff gauge invariants}

 Let us start with the NSDiff transformation $\delta e_{\mu\nu}=\partial_\nu\epsilon_\mu$. We can use $D$ out of those $D^2$ equations and find, with help of (\ref{decome}), \be\alpha=\delta\left[ \frac{\partial_{j}e_{0j}}{\nabla^2} \right]\, ; \qquad \beta^{t}_j=\delta\left[ \frac{\theta_{jk}\partial_ie_{ki}}{\nabla^2} \right] \, ; \qquad \lambda=\delta\left[ \frac{\partial_{i}\partial_{j}e_{ij}}{\nabla^4} \right]\label{helicitys2}\,.\ee

 \no The remaining equations lead to \be \delta {\cal I}_{00}\equiv\delta\left[ e_{00}- \frac{\partial_{j}\dot e_{0j}}{\nabla^2} \right]=0\, ;\qquad  \delta \bar{\mathcal I}^t_{j} \equiv\delta[\theta_{ij}e_{0i}]=0\label{e0j}\,;\ee
\be \delta\tilde {\bar {\mathcal I}}_{j} \equiv \delta\left[ e_{j0}-\frac{\partial_i \dot e_{ji}}{\nabla^2}\right]=0\, ;\qquad  \delta\bar {\mathcal I}_{ij} \equiv \delta\left[\theta_{jk}e_{ik}\right]=0\label{eij}\,.\ee

The NSDiff symmetry also contains antisymmetric transverse shifts $\delta_\Lambda e_{\mu\nu}=\Lambda_{[\mu\nu]}^{T}$, therefore we need to guarantee this symmetry for the invariant NSDiff basis. The following decomposition satisfies $\p^{\mu}\Lambda_{\mu\nu}^T=0$,

\be \Lambda_{0j}^T=\lambda_j^t \, ;\quad \Lambda_{ij}^T=\lambda^{tt}_{ij}+\frac{1}{\nabla^2}(\partial_{i}\dot\lambda^t_j-\partial_j\dot\lambda_{i}^t)\, . \ee 

\no Besides the obvious invariance  $\delta_\Lambda {\cal I}_{00}=0$, we have

\be  \delta_\Lambda \bar{\mathcal I}^t_{j} =\lambda_j^t \, ; \qquad \delta_\Lambda \tilde{\bar{\mathcal I}}_{j} =-\frac{\Box}{\nabla^2}\lambda_j^t \, ; \qquad \delta_\Lambda \bar{\mathcal I}_{ij}=\lambda_{ij}^{tt}+\frac{\partial_i\dot\lambda_j^t}{\nabla^2}\label{eijshift}\,.\ee 

\no Therefore, altogether with ${\cal I}_{00}$, we have the following invariants under (\ref{dnstd})

\be\delta_\Lambda\mathcal I_{j}\equiv\delta_\Lambda\left(\frac{{\Box}}{\nabla^2}\bar{\mathcal{I}}_j^t+\tilde{\bar{\mathcal{I}}}_j\right)=0\,;\qquad \delta_\Lambda\mathcal{I}_{ij}\equiv\delta_\Lambda \left(\bar{\mathcal{I}}_{ij}+{\bar{\mathcal{I}}}_{ji}-\frac{\partial_i\dot{ \bar{\mathcal{I}}}_{j}^t}{\nabla^2}-\frac{\partial_j\dot {\bar{\mathcal{I}}}_{i}^t}{\nabla^2}\right)=0\, .\ee 

\no Noticing the identity $\p^i\p^j\mathcal{I}_{ij}=0$, it becomes clear that the number of NSDiff invariants $\{\mathcal{I}_{ij},\mathcal I_{j},{\cal I}_{00}\}$, i.e.,  $(D-1)D/2 -1 + (D-1) +1 = D(D+1)/2 -1$  matches  (\ref{ni}), i.e.,  

\be N_I^{NSD}= N(e_{\mu\nu})- N(\eps_{\mu}) - N(\Lambda^T_{\mu\nu}) =   D^2- D - \left[\frac{D(D-1)}2-(D-1)\right] = \frac{D(D+1)}2-1 \, . \label{ninsdiff} \ee

It is convenient to redefine the NSDiff basis. Starting with $\mathcal{I}_j$, we will split it into its transverse ($\mathcal{I}_{j}^{t}\equiv\theta_{ij}\mathcal I_i$) and longitudinal ($\mathcal I\equiv\partial_j\mathcal I_{j}$) components\be \mathcal{I}_j^t=\theta_{ij}\left(\Box e_{0i}+\nabla^2e_{i0}-\partial_k\dot e_{ik}\right)/\nabla^2\ ,\qquad \mathcal{I}={\partial_j e_{j0}}-{\omega_{ij}\dot e_{ij}}.\ee The invariants $\mathcal{I}_{ij}$ can be decomposed in a irreducible form consistent with $\p^i\p^j \mathcal I_{ij} =0 $: \be \mathcal I_{ij}=2\,I_{ij}^{tt}+\frac{\partial_i\tilde{\mathcal I}_{j}^t}{\nabla^2}+\frac{\partial_j\tilde{\mathcal I}_{i}^t}{\nabla^2}+2\frac{\theta_{ij}}{(D-2)} I \ ,\ee with \be \tilde{\mathcal I}_{j}^t={\theta_{jk}}(\partial_i e_{ik}-\dot e_{0k})\, , \qquad I=\theta_{ij}h_{ij}\ ,\qquad  I_{ij}^{tt}=\left[\theta_{ik}\theta_{jl}-\frac{\theta_{ij}\theta_{kl}}{(D-2)}\right]h_{kl}\ .\ee 

\no In summary, we have the following basis of NSDiff invariants

\be I_{A}^{NSD}=\{I_{ij}^{tt},\mathcal{I}_{j}^{t},\mathcal{\tilde{I}}_j^{t},{\cal I}_{00},\mathcal{I}, I\} \quad . \label{nsdinv} \ee

\no The NSDiff theory that appears in (\ref{lc}) becomes:
\bea \mathcal{L}_{NSD}(c)&=&I_{ij}^{tt}\frac{\Box}{2}I_{ij}^{tt}-I_{j}^{t}\,\nabla^2\, I_{j}^{t}-\frac{(D-1)}{2(D-2)}I\, \Box \, I + \frac{I\, \nabla^2 \, I_W}{(D-2)} \nn\\
&-& c\, \left\lbrack \mathcal I_{00} \, \nabla^2 \, \mathcal I_{00}  - 2\,  \mathcal I_{00}  \nabla^2 \, I  + I\, \Box \, I +2\, I \, \dot{\mathcal I} + {\mathcal I}^2  - (\tilde{\mathcal I}_j^t)^2 \right\rbrack \, , \label{ssd} \eea

\no The first line of (\ref{ssd}) is the linearized EH theory. The invariants $\{I_{ij}^{tt},I_j^t,I,I_W\}$ are the same symmetric Diff invariants of the last section, in particular, $I_W=(D-2)\, I_{00} + \Box\, I/\nabla^2$. In terms of the NSDiff basis
(\ref{nsdinv}) we have $I_{00} = {\cal I}_{00}- \dot{{\cal I}}/\nabla^2$ and $I_j^t = ({\cal I}_j^t - \dot{\tilde{{\cal I}}}^t_j/\nabla^2)/2$. 

At $c=0$ we only have a massless spin-2 particle. Assuming $c\ne 0$, in order to diagonalize the $SO(D-1)$ scalar sector, we can redefine some invariants,

\be {\cal \overline{I}}=\mathcal{I}-\frac{(1+2c)}{2c}\dot{I}\quad ;\quad {\cal\overline{ I}}_{00}={\cal I}_{00}-\frac{(1+2c)}{2c}I\, ,\ee 

\no consequently (\ref{ssd}) becomes

\be \mathcal{L}_{NSD}(c)=I_{ij}^{tt}\frac{\Box}{2}I_{ij}^{tt}+\frac{(D-1)(c+c_D)}{4c\,(D-2)}I\Box I-I_{j}^{t}\,\nabla^2\,I_{j}^{t}+c(\tilde{\mathcal I}_{j}^{t})^2-c\,{\cal \overline{I}}^2-c\,{\cal\overline{ I}}_{00}\,\nabla^2\,{\cal\overline{ I}}_{00}\label{lci}\, ,\ee

\no which confirms the particle content previously anticipated, namely, a physical massless spin-2 particle ($I_{ij}^{tt}$) and a spin-0 particle ($I$) which is physical if either $c>0$ or $c<-c_D$. The remaining invariants are nonpropagating. Thus, from (\ref{lc}), we conclude that $\mathcal{L}_{NST}(c,B)$ is an SST physical theory if $B>0$ and either $c>0$ or $c<-c_D$. We only need to guarantee the independence of the gauge invariants as we have done in the symmetric case.

The NSTDiff symmetry has one less gauge parameter and one more gauge invariant when compared to NSDiff due to the transversality of the gauge parameter. We can use the trace $h$ as the additional gauge invariant. To show the linear independence of the invariants, again we resort to the helicity variables decomposition\be e_{00}=\rho;\ \ e_{0j}=\gamma_j^++\partial_j\Gamma^+;\ \  e_{j0}=\gamma_j^-+\partial_j\Gamma^-;\ee
\be e_{ij}=\Psi_{ij}^{tt}+\beta_{ij}^{t}+\partial_{i}\Phi_j^++\partial_j\Phi_{i}^{-}+\nabla^2(\theta_{ij}\Lambda+\omega_{ij}\psi),\ee 

\no where $(\Psi_{ij}^{tt},\beta_{ij}^{t})=(\Psi_{ji}^{tt},-\beta_{ji}^{t})$ and $\delta_{ij}\Psi_{ij}^{tt}=\partial_i \Psi_{ij}^{tt}=\partial_i\beta_{ij}^{t}=\partial_i \gamma_i^+=\partial_i\gamma_i^-=\partial_i \Phi_i^+=\partial_i\Phi_i^-=0$. We only need to focus on the $SO(D-1)$ scalars and vectors: \be {\cal {\overline I}}_{00}=\rho-\dot \Gamma^+-\frac{(1+2c)(D-2)}{2c}\nabla^2\Lambda\, ;\qquad \overline{{\mathcal I}}=\Gamma^--\dot\psi-\frac{(1+2c)(D-2)}{2c}\nabla^2\dot\Lambda\, ;\ee \be I=(D-2)\nabla^2\Lambda \, ;  \qquad h=-\rho+(D-2)\nabla^2\Lambda+\nabla^2\psi\label{h51} \ee \be I_{j}^{t}=\frac{1}{2}(\gamma_j^++\gamma_{j}^--\dot\Phi_j^+-\dot\Phi_j^-)\, ;\qquad \tilde {\mathcal I_{j}}^{t}=\nabla^2\Phi_j^+-\dot \gamma_j^+\, .\label{vectors}\ee

In the scalar sector, we can write the following relation between helicity variables and the gauge invariant basis\be \begin{bmatrix}
\bar {\cal I}_{00} \\
I \\
\bar I \\
h
\end{bmatrix}=\mathbb{M}_{NTD}\begin{bmatrix}
\rho \\
\Lambda \\
\Gamma^- \\
\psi
\end{bmatrix}-\begin{bmatrix}
\dot \Gamma^+ \\
0 \\
0 \\
0
\end{bmatrix} \, ,\ee where \be \mathbb{M}_{NTD}\equiv\begin{bmatrix}
1 & -\frac{(1+2c)(D-2)\nabla^2}{2c} & 0 & 0\\
0 & (D-2)\nabla^2 & 0 & 0\\
0 & -\frac{(1+2c)(D-2)\nabla^2\del_0}{2c} & 1 & -\del_0\\
-1 & (D-2)\nabla^2 & 0 & \nabla^2
\end{bmatrix}\ee Since  $det[\mathbb{M}_{NTD}]=(D-2)\nabla^4\, $ the four scalar invariants are independent quantities. Similarly, in the $SO(D-1)$ vector sector, the determinant of the relevant matrix that takes us from $\{\gamma_j^-,\Phi_j^+\}$ to $\{I_j^t,\tilde {\mathcal I_{j}}^{t}\}$ is given by $\nabla^2$, so we have again a nonsingular change of variables.
In the appendix, we alternatively confirm the particle content of the NSTDiff model (\ref{lc}) via an auxiliary vector field without using Bardeen variables. 


Now, when going from the NSDiff to the NSWTDiff basis, we recall that  $\bar {\cal I}_{00}$, $\bar {\cal I}$ and $\mathcal I$ are the only Weyl noninvariant Bardeen variables. Using the transformation $h_{\mu\nu} \to \bar{h}_{\mu\nu}(h) = h_{\mu\nu} - \eta_{\mu\nu}\, h/D$ to obtain the NSWTDiff invariants, we reach

\be \bar {\cal I}_{00}^{WT}=\left[\frac{2c-(D-2)}{2cD}\right]e_{00}+\left[\frac{(D-2)+2c(D-1)}{2cD}\right]e_{jj}-\frac{\partial_j\dot e_{0j}}{\nabla^2}-\frac{(1+2c)}{2c}\theta_{ij}e_{ij}\, ;\ee \be \bar {\cal I}_{WT}=\frac{\partial_j e_{j0}}{\nabla^2}+\left[\frac{(D-2)-2c}{2cD}\right]\frac{\dot e_{jj}}{\nabla^2}-\left[\frac{2c(D-1)+(D-2)}{2cD}\right]+\frac{\dot e_{00}}{\nabla^2}-\frac{\theta_{ij}\dot e_{ij}}{2c\nabla^2}\label{Iwt}\, ;\ee

\be I_{WT}=\theta_{ij}e_{ij}-\frac{(D-2)h}{D}\ ,\ee 

\no Therefore, the scalar NSWTDiff invariants appearing in the ${\cal L}_{NSWTD}(h)={\cal L}_{NSD}(\bar{h}(h))$ are given by $\{{\cal I}_{00}^{WT}, {\cal \overline I}_{WT},  I_{WT}\}$ and as expected from (\ref{ni}), we have $N_I^{NSDiff}=N_I^{NSWTDiff}$. Since they share the same vector sector, we only need to draw our attention to the independence of $h$, given in (\ref{h51}) and the scalar sector, namely,

\be \bar {\cal I}_{00}^{WT}=\left[\frac{2c-(D-2)}{2cD}\right]\rho+\left[\frac{(c+1)(D-2)}{cD}\right]\nabla^2\Lambda+\left[\frac{(D-2)+2c(D-1)}{2cD}\right]\nabla^2\psi\, ; \ee \be \bar {\cal I}_{WT}=-\left[\frac{2c(D-1)+(D-2)}{2cD}\right]\dot\rho-\left[\frac{(D-2)(c+1)}{cD}\right]\dot\Lambda+\Gamma^-+\left[\frac{(D-2)-2c}{2cD}\right]\dot\psi\, ;\ee
\be I_{WT}=\frac{(D-2)}{D}\left[\frac{\rho}{\nabla^2}+2\Lambda-\psi\right]\, .\ee In matrix form,

\be \begin{bmatrix}
\bar {\cal I}_{00}^{WT} \\
\bar {\cal I}_{WT} \\
I_{WT} \\
h 
\end{bmatrix} =\mathbb{M}_{NSW}\begin{bmatrix}
\rho \\
\Lambda \\
\Gamma^- \\
\psi 
\end{bmatrix} -\begin{bmatrix}
\dot \Gamma^+ \\
0 \\
0 \\
0 
\end{bmatrix},\ee where \be \mathbb{M}_{NSW}\equiv\begin{bmatrix}
\left[\frac{2c-(D-2)}{2cD}\right] & \frac{(c+1)(D-2)}{cD}\nabla^2 & 0 & \left[\frac{(D-2)+2c(D-1)}{2cD}\right]\nabla^2 \\
-\left[\frac{2c(D-1)+(D-2)}{2cD}\right]\del_0 & -\left[\frac{(D-2)(c+1)}{cD}\right]\del_0 & 1 & \left[\frac{(D-2)-2c}{2cD}\right]\partial_0 \\
\frac{(D-2)}{D\nabla^2} & 2\frac{(D-2)}{D} & 0 & -\frac{(D-2)}{D} \\
-1 & (D-2)\nabla^2 & 0 & \nabla^2 
\end{bmatrix}\, ,\ee Since $\det (\mathbb{M}_{NSW})=-(D-2)\nabla^2$, the quantities $\{{\cal I}_{00}^{WT}, {\cal \overline I}_{WT},  I_{WT}\}$ form a NSWTDiff independent set of invariants.

 \section{Coupling to external sources}
 
 Now we couple the SST model (\ref{lc}) to an external nonsymmetric tensor 
 $T_{\mu\nu} \ne T_{\nu\mu}$ and single out the contributions of the two scalars to the two-point amplitude as compared to the usual linearized EH theory. We start with
 
 \be {\cal L} (c,B,T) = \frac{1}{2} h^{\alpha\beta}\, \Box \,  h_{\alpha\beta}+(\del^\mu h_{\mu\beta})^2+ \bar{A}\, h\, \del^{\mu}\del^{\nu} h_{\mu\nu}-\frac{\bar{B}}{2} h\,  \Box h \, + c \, (\p^{\mu}e_{\mu\nu})^2 + e_{\mu\nu}\, T^{\mu\nu} \, . \label{cbt} \ee
 
 \no where $(\bar{A},\bar{B})= (1+2\, c, 1+ 2\, c - B)$. The invariance of the source term $\delta \int d^Dx \, e_{\mu\nu}\, T^{\mu\nu} =0$  under the NSTDiff transformations (\ref{dnstd}) leads to the following constraints on the nonsymmetric source,
 
 \bea \p^{\nu} T_{\mu\nu} &=& \p_{\nu}\, J  \, , \label{nc} \\ 
  T_{\mu\nu} &=& T_{(\mu\nu)} + \p_{\mu} T_{\nu}^T - \p_{\nu} T_{\mu}^T  \quad . \label{dcp} \eea
 
 \no where $T_{(\mu\nu)}=(T_{\mu\nu}+T_{\nu\mu})/2 $ while $J$ and $T_{\mu}^T$ are arbitrary functions except for $\p^{\mu} T_{\mu}^T=0$. If we use an auxiliary vector field $v^{\mu}$ and (\ref{dcp})  we can rewrite (\ref{cbt}) as,
 
 \bea {\cal L} (c,B,T) &=& {\cal L}_{TD}(\bar{A},\bar{B}) + h_{\mu\nu} T^{(\mu\nu)} + 2\, c\, v^{\mu} \p^{\alpha}h_{\alpha\mu} - c\, v^{\mu}v_{\mu}\nn\\  &+& 2\, B^{\mu\nu}\, (\partial_\mu T_{\nu}^T- c\, \partial_\mu v_{\nu})  \, . \label{fmn} \eea
 
 \no Integrating over $B_{\mu\nu}$ we obtain a vanishing ``field strength'' leading to a functional constraint establishing $c\, v_{\mu} =T^T_{\mu} - \p_{\mu} \, \psi$ where $\psi$ is  an arbitrary scalar field. Integrating over $\psi$ we obtain a nonlocal theory:

 \be {\cal L} (c,B,T) = {\cal L}_{TD}(\bar{A},\bar{B})  - c\, \p^{\mu}\p^{\nu}h_{\mu\nu}\, \frac 1{\Box}\, \p^{\mu}\p^{\nu}h_{\mu\nu} + h_{\mu\nu}\, S^{\mu\nu} - \frac 1c T_T^{\mu} T^T_{\mu} \quad , \label{nl1} \ee
 
 \no The tensor $S_{\mu\nu} \equiv T_{(\mu\nu)} - \p_{\mu} T_{\nu} ^T - \p_{\nu} T_{\mu} ^T$ is a kind of symmetric version of $T_{\mu\nu}$ since it obeys
 
 \bea S &=& T \quad  , \label{st} \\
  \p^{\mu} S_{\mu\nu}&=&\p^{\mu} T_{\nu\mu} = \,\,\p_{\nu}\, J \quad . \label{dels} \eea

 \no where $(S,T)=(\eta^{\mu\nu}S_{\mu\nu},\eta^{\mu\nu}T_{\mu\nu}) $. From (\ref{nl1}) and  (\ref{dels}) it becomes clear that the integration over the antisymmetric components $B_{\mu\nu}$ has led us to a nonlocal TDiff (NLTDiff) model. Moreover, it is natural to introduce $\tilde{S}_{\mu\nu}=S_{\mu\nu}- \eta_{\mu\nu}\, J$ which is  conserved  $\p^{\mu}\tilde{S}_{\mu\nu}=0$ and satisfies $\tilde{S} = T- \, D\, J$. It opens the way to split the NLTDiff model (\ref{nl1})
 into two independent parts involving a nonlocal diffeomorphism invariant model with conserved source (NLDiff) plus a trace dependent term coupled to the conservation breaking function $J$. Namely, 
 
 \be {\cal L} (c,B,T) = {\cal L}^{NLD}(h_{\mu\nu},\tilde{S}_{\mu\nu}) + \frac B2 \, h\, \Box \, h \, + h\, J - \frac 1c T_T^{\mu} T^T_{\mu} \quad . \label{nl2} \ee
 
 \no where,

  \be  {\cal L}^{NLD} =  \frac{1}{2} h^{\alpha\beta}\, \Box \,  h_{\alpha\beta}+(\del^\mu h_{\mu\beta})^2+ (2\, c+1)( h\, \del^{\mu}\del^{\nu} h_{\mu\nu}- \frac 12 h\,  \Box h ) \,  - c\, \p^{\mu}\p^{\nu}h_{\mu\nu}\, \frac 1{\Box}\, \p^{\mu}\p^{\nu}h_{\mu\nu} + h_{\mu\nu}\, \tilde{S}^{\mu\nu} \quad , \label{nld} \ee
  
 It is easy to check that (\ref{nld}) is invariant under arbitrary diffeomorphisms $\delta h_{\mu\nu} = \p_{\mu} \eps_{\nu} + \p_{\nu}\eps_{\mu}$, since the c-dependent terms correspond to $c\, R_L(1/\Box)R_L$, with the linearized scalar curvature $R_L= \p^{\mu}\p^{\nu}h_{\mu\nu}- \Box \, h$, while the remaining c-independent terms correspond to the linearized EH theory. Due to the diff symmetry (including the source term) we know that $ {\cal L}^{NLD}= {\cal L}^{NLD}(I_K^{Diff}) $. Therefore, we can make the functional integral over the trace independently. The final result for the effective action (two point amplitude) in momentum space is given by 
 
 \be {\cal A}_2(p) = \frac{i}{2\, p^2} \left\lbrack \vert \tilde{S}_{\mu\nu}\vert^2 - \frac{\vert \tilde{S} \vert^2 }{D-2} +\frac{c\,\vert \tilde{S} \vert^2}{2(D-1)(D-2)(c+c_D)} + \frac{\vert J \vert^2 }{B}\right\rbrack   - \frac ic  (T^T_{\mu})^* (T_T^{\mu})  \quad . \label{A2} \ee

 \no where we recall that $\tilde{S}_{\mu\nu}= S_{\mu\nu}- \eta_{\mu\nu}\, J$ is conserved $\p^{\mu}\tilde{S}_{\mu\nu}=0$.
 
 The first two terms of (\ref{A2}) is the usual EH amplitude (physical massless spin-2) while the third one is the contribution of the spin-0 field present in (\ref{nld}) or equivalently in ${\cal L}_{NSD}(c)$ before integration over $B_{\mu\nu}$. In particular, the imaginary part of the residue at the pole $p^2=0$ is positive (physical scalar) if either $c>0$ or $c<-c_D$ which confirms our expectations. The  term $ \vert J \vert^2 /{B} $ represents the second massless scalar field stemming from the trace $h$. It is physical if $B>0$ as expected. The last term has no particle interpretation (no poles). It is a contact term, a relic of the antisymmetric components of $e_{\mu\nu}$. The contribution of both scalars for physical phenomena as, for instance, the gravitational lenses effect is always positive and cumulative, increasing the deviation angle as compared to the pure EH contribution. We can only have negative contributions if one or both scalars are ghostly, similarly to the pure TDiff model as commented in \cite{rr}.
 
 Finally, suppressing the sources, although there is in general no unique non linear completion of linearized theories, the NLDiff  model (\ref{nl2}) obtained after integration over the antisymmetric components of $e_{\mu\nu}$ may be considered as a linearized version  about the flat Minkowiski space  of the following nonlocal model:
 
 \be {\cal L}_g = \sqrt{-g}\left\lbrack C_1(-g) R + C_2(-g)\, R\, \frac 1{\Box} \, R  + C_3(-g)\, g^{\mu\nu} \p_{\mu}\, g\, \p_{\nu}\, g \right\rbrack\quad . \label{cov} \ee
 
 \no where $R$ is the usual scalar curvature while $C_j(-g)\, \, , j=1,2,3 $ are  functions of the metric determinant $g$ analytic at $g=-1$ such that: $C_1(1)>0$, $C_1^{\prime}(1)=0$, $C_2(1)=c$ and  $C_3(1)=-B\, C_1(1)/4$. This can be checked\footnote{Recall that (70) at $B=0=c$, without sources, corresponds to twice the linearized Einstein-Hilbert theory $\sqrt{-g}R$ about the flat Minkowiski space. Moreover, $g=-1-h +{\cal O}(h^2)$ and since the scalar curvature is of first order in $h_{\mu\nu}$, the non local term in (\ref{cov}) becomes the non local term of (71) at quadratic order.} by performing a weak field expansion according to  $g_{\mu\nu} = \eta_{\mu\nu} + h_{\mu\nu}/\sqrt{C_1(1)})$. The action corresponding to (\ref{cov}) is invariant under volume preserving diffeomorphisms. Infinitesimally it amounts to $x^{\mu} \to x^{\mu} + \eps^{\mu}$ with $\nabla_{\mu}\eps^{\mu}=0$. The metric determinant behaves as a scalar field under such transformations, see comments in \cite{blas} and \cite{bsz}. In the latter reference, a sub case of (\ref{cov}) where the nonlocal term on the curvature is absent, i.e. an ST model with a spin-2 and only one scalar mode, has been investigated phenomenologically. It might be interesting to ``localize'' the nonlinear SST theory (\ref{cov}) and search for phenomenological effects. This is among our goals.

%
%
%

 \section{Conclusion}
 
Here we have investigated the most general Lorentz invariant massless free model of second order in derivatives, described by a rank-2 tensor with both symmetric and antisymmetric parts and invariant under transverse diffeomorphisms ($\p^{\mu}\eps^T_{\mu}=0$). By demanding a  healthy massless spin-2 particle in the spectrum, we have found three classes of models given in (\ref{labct}), (\ref{lc}) and (\ref{lnswtd}), each parametrized by two real parameters. They are displayed in Table 1.   Except for special points in the parameter's space where the spectrum is reduced to previously known models in the literature, they all feature two massless scalars (spin-0) and one massless spin-2 (tensor) particle in their spectrum which we call an SST model. There is always a continuous family of SST models such that all three particles are physical. 

Coupling the SST models (\ref{lc}) to a general source compatible with the  NSTDiff symmetry (\ref{dnstd}) revealed that the contribution of the two scalars is independent but additive, see (\ref{A2}), confirming the particle content predicted by the improved gauge invariant method that we have used. Furthermore, as pointed out in \cite{rr} for the case of the pure (symmetric) TDiff model with only one scalar field alongside the graviton, the scalar contribution to the deflection angle of the light beams by the sun (gravitational lens effect) is always positive for each scalar. This is a consequence of the requirement of unitarity for both spin-0 particles. The last term in (\ref{A2}) is a contact term and comes entirely from the  antisymmetric part of the tensor field. 

A critical issue  is the physical equivalence of different points in the parameters space of the SST model. Recently, we have shown that all physical ($f_D(a,b)>0$) TDiff models ${\cal L}_{TD}(a,b)$, see (\ref{sab}), are equivalent to the simplest case $a=b=0$, a spin-2 version of the higher spin Maxwell-like models of \cite{cf}. We have shown that the two point amplitude
of a $(a,b)$ model is equivalent to the amplitude of the $(0,0)$ model once we redefine the  source coupling accordingly. We are currently investigating this issue regarding the models (\ref{lc})  and (\ref{lnswtd}).

An important direction to follow is to search for non linear completions of the consistent free models of table 1. As a first step, we are now using (work in progress) the Noether iterative method, see, e.g., \cite{zino3}, in order to investigate the construction of consistent interacting cubic vertices along with extensions of the linearized gauge symmetry, similar in spirit to the Gupta program which starts with a spin-2 linearized model and aims to arrive at a full non linear gravitational model via an iterative procedure, see \cite{gupta} and \cite{deserr}, with the difference that now we are dealing with a rank-2 tensor with an antisymmetric part. On the other hand, one can think of integrating out the antisymmetric part of the tensor as in the case of the SST models (\ref{lc}). So we have obtained a nonlocal model whose natural nonlinear completion is given by the covariant, under volume preserving diffeomorphisms, theory (\ref{cov}) where the restrictions on the functions $C_j(-g)\, , j=1,2,3$, are given right after (\ref{cov}). Although turning a nonlocal field theory into a local one  via auxiliary fields is a nontrivial issue, see for instance \cite{bdfm},  one might think of replacing $C_3(-g)\sqrt{-g} R(1/\Box)R $ by $ C_3(-g)\sqrt{-g}(2\,R\, \psi +\psi \, \Box \, \psi)$ in (\ref{cov}) and investigating the phenomenological consequences of a generalization of the gravitational scale invariant theory studied in \cite{bsz} which corresponds to $ C_3(-g)=0$. 

Another application of our work concerns the general teleparallel quadratic gravity investigated  in \cite{gtg} in $D=4$. It corresponds to a diffeomorphism invariant generalized quadratic form on nonmetricity and torsion scalars with vanishing curvature. The fundamental fields are the metric $g_{\mu\nu}$ and $\Lambda_{\mu}^{\,\,\alpha}$, an element of $GL(4,\mathbb{R})$.  Expanding about Minkowski space one ends up with a quadratic theory of two symmetric and one antisymmetric rank-2 tensor. Besides the diffeomorphisms, one has to add extra gauge symmetries to get rid of ghosts. The authors of \cite{gtg} have considered a maximum of five degrees of freedom corresponding to two spin-2 particles and one spin-0 coming from the two-form field, however based on our findings here, we believe that requiring TDiff as the extra symmetry we may even have, at linearized level, up to six healthy degrees of freedom, corresponding to two spin-2 particles and two scalars, this is under investigation.

We remark that the gauge invariant approach adopted here,  rooted in Bardeen variables \cite{Jaccard},  makes the identification of the physical content of the whole model straightforward. Although Lorentz covariance is not manifest, it can be shown that  the final theory is Lorentz invariant. We believe that the method may be especially useful in higher rank tensor theories where lower spin (traces) components complicate conventional analyses. Moreover, recently one has shown \cite{bddm} the equivalence of different points in the parameters space of some spin-2 and spin-3 models invariant under transverse differomorphisms. We believe that by means of redefinitions of gauge invariants and sources we can achieve a simpler proof of a similar equivalence regarding the  parametric families of models
in table 1.

Finally, It would be interesting also to figure out how the reduction in the number of effective field components stemming from the use of gauge invariants is connected  with the absence of redundant field components in some first order wave equations approach to higher spin particles as in \cite{simulik}.

\section{Acknowledgements}

The work of D.D. is partially supported by CNPq  (grant 306380/2017-0), LGMR is supported by CAPES. 

\section{Appendix - NSTDiff content via auxiliary field}\renewcommand{\theequation}{A-\arabic{equation}}\setcounter{equation}{0}

Notice that (\ref{lc}) can be written as a symmetric TDiff model
(\ref{sab}) and one additional term,\be \mathcal{L}_{NST}(c,B)=\mathcal{L}_{TD}(\bar a,\bar b)+c\,(\del_\mu e^{\mu\nu})^2 \label{everyform}\ee where $(\bar a, \bar b)=(2c+1,2c+1-B)$. An auxiliary vector field $v_{\mu}$ can be used to rewrite (\ref{everyform}) without changing its physical content, \be S_{NST}[h_{\mu\nu},B_{\mu\nu},v_\mu]=\int\, d^Dx \left[\mathcal{L}_{TD}(\bar a,\bar b)-c\,v_\nu v^\nu+2c\,v_\nu(\del_\mu h^{\mu\nu}+\del_\mu B^{\mu\nu})\ \right]\, .\label{snst} \ee  

\no Integrating over $B_{\mu\nu}$, which appears linearly in (\ref{snst}), we obtain the following constraint and its general solution in term of a scalar field \be \del_\mu v_\nu-\del_\nu v_\mu=0\Rightarrow v_\mu=\del_\mu\psi \label{constraint}\, ,\ee

\no So (\ref{everyform}) is equivalent to \be \mathcal{L}_{NST}(c,B)=\mathcal{L}_{TD}(\bar a, \bar b)-c\,\del_\mu\psi\,(\del^\mu\psi-2\del_\nu\,h^{\mu\nu}) \label{lap}.\ee 

\no It is convenient to define

\be q \equiv \frac{B\, D^2}{(D-2)(D-1)} \quad ; \quad p \equiv 2\,c\,D + (D-2) \quad . \label{qp}\ee

\no Assuming $p \ne 0$, the redefinition $h_{\mu\nu}\rightarrow \tilde{h}_{\mu\nu}+ (2c/p)\,\eta_{\mu\nu}\psi$ leads to

\be \mathcal{L}_{NST}(c,B)=\mathcal{L}_{TD}(\bar a, \bar b)+K\,\psi\,\Box\,\psi + L\, \tilde{h}\,\Box\,\psi \label{LK} \ee 

\no where 

\be K\equiv\frac{2(D-2)(D-1)}{p^2}\,  \left[c(1+q)+c_D\right]\, c  \quad ; \quad L\equiv\frac{2c}p(1+\bar a-\bar bD)\quad  .\ee 

\no Further assuming $K \ne 0$, after  $\psi \to \phi - L\, \tilde{h}/(2K)$, we end up with a scalar field decoupled from a TDiff model,

\be \mathcal{L}_{NST}(c,B)=\mathcal{L}_{TD}(\bar a, b^*)+K\,\phi\,\Box\,\phi \label{A8} \ee 

\no where $ b^* = \bar b + L^2/(2\, K)$. Consequently, see (\ref{fd}), we have 
a healthy SST model whenever,

\be K >0 \quad {\rm and} \quad f_D(\bar a, b^*) = \frac{p^2B(c+c_D)}{(D-2)\left[c(1+q)+c_D\right]} > 0 \quad . \label{conds}\ee

\no \no  {\bf Case I}: $\mathbf{B<0}$

In this case we have $q+1 < 1$. There are three sub cases. If $q+1=0$ the condition $K>0$ leads to $c>0$ then it follows that $f_D<0$ which violates the second condition (\ref{conds}). If $0<q+1<1$, the condition $K>0$ leads to either $c>0$ or $c<-c_D/(q+1)$. The case $c>0$ must be ruled out, it leads to $f_D<0$ again. If $c<-c_D/(q+1)$ we have $c < - c_D$ due to $0<q+1<1$, it leads once more to  $f_D<0$ and must be dicarded too. We are left with  $q+1 < 0$. In this case $K>0$ requires $0<c<-c_D/(q+1)$, since $f_D>0$ demands now $c>-c_D/(q+1)$ or $c<-c_D$, so there is no solution. 

\no  {\bf Case II}: $\mathbf{B>0}$

Now we have $q+1 > 1$ and $K>0$ requires either $c>0$ or $c<-c_D/(q+1)$. On the other hand, $f_D >0$ demands $c>-c_D/(q+1)$ or $c<-c_D$. So both conditions (\ref{conds}) imply either $c>0$ or $c<-c_D$ in full agreement with our much simpler analysis based on the gauge invariants shortcut. 

Finally, concerning the singular point $p=0$, where
we have 

\be (\bar a, \bar b, c) = (\frac 2D, \frac 2D -B,\frac{(2-D)}{2D}) \equiv (\bar a_0, \bar b_0, c_0)\quad . \label{abc} \ee

\no If we try to decouple $\psi$ from $h_{\mu\nu}$ as in (\ref{lap})
via $h_{\mu\nu} \to \tilde{h}_{\mu\nu} + k \, \eta_{\mu\nu} \, \psi$
we find

\be \mathcal{L}_{NST}(c_0,B)=\mathcal{L}_{TD}(a_0,b_0)+K_0\,\psi\,\Box\,\psi + L_0\, \tilde{h}\,\Box\,\psi + 2\, c_0 \, \p^{\mu}\tilde{h}_{\mu\nu}\, \p^{\nu} \psi\label{LK0} \ee 

\no where 

\be K_0 = \frac{k^2}2 (B\, D^2 + 2 - D) + c_0 (1-2\, k) \quad ; \quad L_0 = \frac k{2\,D}(B\, D^2 + 2 - D) \quad . \label{kl0} \ee

\no Because of the last term in (\ref{LK0}) there is no decoupling solution for $k$. The auxiliary field trick fails in this case while the gauge invariants method still holds. We could have integrated over $\psi$ already in (\ref{lap}) leading to a spacetime nonlocal TDiff model. However, in order not to lose part of the content, we should have introduced sources from the start before integration as we have done in section 4.

\end{document}